\documentstyle[aps,twocolumn,prl,epsf]{revtex}

\begin{document}
\draft

\twocolumn[\hsize\textwidth\columnwidth\hsize\csname @twocolumnfalse\endcsname
\title{A scenario for the electronic state in the manganase perovskites:\\ 
the orbital correlated metal}

\author{Marcelo J. Rozenberg
}

\address{Institut Laue--Langevin, BP156, 38042 Grenoble, France
}

\date{\today}
\maketitle
\widetext
\begin{abstract}
\noindent

We argue that, at low temperatures and well into the ferromagnetic phase,
the physics of the manganase perovskites
may be characterized by a correlated metallic
state near a metal insulator transition 
where the orbital degrees of freedom play a main role. This follows
from the observation that a two-band degenerate Hubbard model 
under a strong magnetic field can be mapped onto a para-orbital single band
model. We solve the model numerically using the quantum Monte Carlo technique
within 
a dynamical mean field theory which is exact in the limit of large
lattice connectivity.
We argue that the proposed scenario may allow for the qualitative interpretation
of a variety of experiments which 
were also observed in other (early) transition metal oxides.
\end{abstract}

]

\narrowtext

There is a great current interest in transition metal compounds displaying
colossal magnetoresistance (CMR). This effect is a strong dependence of 
resistivity with
the applied magnetic field and is observed experimentally
in compounds such as ${\rm La_{1-x}A_xMnO_3}$, with ${\rm A=Sr, Ca, Pr}$.
{}From the point of view of the electronic structure, these systems have 3
electrons in a  $t_{2g}^3$ band
which due to a strong Hund's rule coupling form a core $3/2-$spin
at each Mn site 
and ${\rm 1-x}$ electrons that go into a quasi two-fold degenerate 
$e_g$ band. For ${\rm x=0}$ the Manganese atoms are (+4) and the compound has
a nominal filling equal to 1.
Since the bands are originated from partially filled d-orbitals, 
on general grounds one expects that correlation 
effects should play an important role in the low energy behavior, which
is indeed
the case in many other transition metal oxides (TMO) with the perovskite
structure \cite{fujimori}. 

The most qualitative aspects of the CMR is explained by the double 
exchange mechanism (DEM) proposed by Zener almost 50 years ago.
More recently, a detail studies of the ferromagnetic Kondo lattice
sowed that while DEM is certainly a main ingredient \cite{furukawa}, 
additional interactions
must be added to obtain a more correct picture \cite{millis1}.
Among those, the dynamical Jahn-Teller effect was studied in detail
\cite{millis2,roder1}
and there is experimental evidence of strong polaronic effects in many 
compounds \cite{roder2}.
In spite of this recent progress, our understanding of the CMR compounds
is still incomplete. In particular at low temperature and within the 
ferromagnetic (FM) phase where there are several experimental observations
that remain unaccounted for:
i) the enhancement of the density of states near the Fermi energy 
with decreasing $T$ observed in  
photoemission \cite{park,ddsarma},
ii) the unusual redistribution of optical spectral weight as function
of the temperature wich occurs in 
the range of the ${\rm eV}$ \cite{okimoto}, iii)
the strong enhancement of
the $A$ coefficient of the $T^2$ term in the resistivity as function 
of the composition
in ${\rm La_{1-x}Sr_xMnO_3}$ \cite{urushibara}, and 
iv) the suppression of the resistivity
with applied pressure \cite{moritomo}. 

These features are not exclusive of the CMR manganese compounds, on the 
contrary, 
they had already been observed in other (early) transition metal oxides
with perovskite structure.
For instance, in ${\rm V_2O_3}$, a redistribution
of  spectral weight in the optical response \cite{v2o3}
and a small enhancement of the density of states near the Fermi energy in the 
photoemission \cite{shin}
is observed as the temperature is lowered within
the paramagnetic metallic phase. 
The ${\rm La_xSr_{1-x}TiO_3}$ system, on the other hand, 
displays a notable increase of 
the $T^2$ coefficient of the resistivity as a function of ${\rm x}$
\cite{tokura}.
Both compounds also show suppression of the resistivity with 
applied pressure \cite{carter,okada}. 
These experiments on early TMO have 
recently received a qualitative
interpretation within a dynamical mean field theory of the Hubbard
model that becomes exact in the limit of 
large dimensions \cite{review,exp}.
The key feature is to realize that the mean field solution of
the Hubbard model predicts a narrow quasiparticle resonance at the Fermi
level when the system is close to a metal insulator transition (MIT).
Thus, the proximity to a MIT provides with a dynamically generated small energy 
scale which allows for the qualitative interpretation
of the unusual behavior observed in the experimental compounds.  

In regard of these similarities between some early TMO and
the manganese perovskites, one is motivated to consider the question whether
a similar underlying mechanism may be responsible for the low energy behavior.
However, an apparent reason that would mean to immediately discard 
this idea is
that the observed phenomena in the CMR compounds occur as they
evolve into a ferromagnetic (FM) state with a 
magnetic moment that saturates close to the classical value. Naively one may 
think that this is 
incompatible with the presence of coherent quasiparticles.
Thus, the goal of this paper is to 
demonstrate that in the parameter regime relevant for the CMR manganates
the system remains close to a MIT with a
dynamically generated
small energy scale as it goes in to the fully polarized
FM state at low temperatures. 
The full Hamiltonian for the CMR compounds has the following form
\cite{millis2,review}:
\begin{eqnarray}
H  = &&  \sum_{<ij>,a,b,\sigma} t_{ij}^{ab}\ \  c_{ia\sigma}^\dagger c_{jb\sigma} 
    - J_H \sum_{i,a,\sigma} \vec{S}_{ci} \cdot \vec{s}_{di} +
\nonumber\\
&& + \ \ H_{J-T} \ \ + \ \ U \sum_{i,a,b,\sigma,\sigma'} 
n_{ia \sigma} n_{ib \sigma'} (1 - \delta_{ab} \delta_{\sigma \sigma'})
\label{hamil}
\end{eqnarray}
where $a,b = 1,2$ are the orbital indexes of the $e_g$ bands and
the local spins 3/2 are described by $\vec{S_c}$ 
The first two terms define the ferromagnetic Kondo lattice 
model and give a realization of the DEM while
the third includes polaronic effects. The role of the local
dynamical correlations due to the last term remains largely unaccounted and
is a main concern in this work. Recent resonant photoemission experiments
indicate that the manganese parent compound is in an intermediate
state between a charge transfer and Mott-Hubbard insulator with
$U \sim 3.5{\rm eV}$ \cite{park}. Another important aspect which
we shall consider explicitely here is the orbital degeneracy, and,
as it turns out, we shall see
that these degrees of freedom will play a crucial role within
the proposed scenario. 

In order to better focus on the role of local repulsion and orbital degeneracy
we shall simplify the Hamiltonian (\ref{hamil}).
Firstly, as we are concerned with the electronic state in the FM phase
we neglect $H_{J-T}$. This term is most relevant  
around and above $T_c$, but moving deep into 
the FM metallic phase its strengh rapidly 
decreases \cite{radaelli,martin,millis2}.
Also, well into the FM phase we can 
assume the local spins 3/2 to be uniform and static.
Therefore, their main effect on the conduction electrons
is to produce an effective local magnetic
field $h_{loc}$ as they become polarized under the action of an implicit
external field.
We emphasize, however, that the real driving force of the FM state
is in the DEM.
The effective local field $h_{loc}$,
being of electronic origin, may be very strong and is estimated
to be about twice the bandwidth \cite{furukawa}.
Finally, on general grounds one expects that 
off-diagonal $t_{ij}^{ab}$ are a fraction of the diagonal hopping, thus 
we set $t_{ij}^{ab}=-t \ \forall a,b$. 
The main role of the 
off-diagonal hopping is to contribute to the stability of the
para-orbital state that we shall encounter latter on. Nevertheless, 
a more important contribution, which renders our choice for $t_{ij}^{ab}$ 
non crucial, is the fact that a
rather large doping $x \gtrsim 0.175$ 
is experimentally necessary for CMR. Thus, disorder is the main
reason for the 
stability of the metallic state against orbital long range order, in analogy
with the N\'eel state in a one band Hubbard model.

We shall consider the model within the dynamical mean field theory
that becomes exact in the limit of large dimensions (or large
lattice connectivity).
We shall assume a semi-circular
density of states $\rho^0(\epsilon)=1/(\pi D) \sqrt{1-(\epsilon/D)^2}$, 
with the half bandwidth $D=1$ and $D=2\sqrt{2}t$.
This $\rho^0$ 
is realized in a Bethe lattice, 
and our choice
is due to both simplicity and the realistic finite bandwidth 
that it provides.
Band structure calculations give $D \sim 1{\rm eV}$ \cite{furukawa} which is
similar to other TMO compounds.

The Hamiltonian can then be mapped onto its associated impurity
problem (a degenerate Anderson impurity in a magnetic field), 
which is supplemented with a self-consistency condition that
enforces the translational invariance \cite{review}. 
The self-consistency
condition reads,
\begin{equation}
[{\cal{G}}^0_{\sigma a}(z)]^{-1}= z + \mu -\sigma h_{loc}
- t^2[G_{\sigma a}(z)+G_{\sigma b}(z)]
\label{selfcons}
\end{equation}
where ${\cal{G}}^0$ and $G$ denote local Green functions,
$\sigma=\pm {1 \over 2}$ the 
spin, $\mu$ the chemical potential and $h_{loc}$ is
the effective local magnetic field due to the $J_H$ coupling of the
conduction electrons with the localized spins $3/2$. 
It is now clear from (\ref{selfcons}) the role of 
the off-diagonal hopping providing
{\it frustration} in orbit space and favoring the stability of 
a para-orbital (orbital disordered) state.

We shall numerically solve this model using the quantum Monte Carlo technique
\cite{fye,note0}.
To demonstrate that the model remains near a MIT
as it goes into the FM phase we shall compute
the mass renormalization $m^*/m$. In the limit of
large dimensions the self-energy $\Sigma={{\cal{G}}^0}^{-1}-G^{-1}$ 
is local, thus, 
$m^* / m= 1- {\partial{\Sigma}}/{\partial{\omega}}$.
Since our results are obtained at low but finite temperature we shall estimate
this value using $m^* / m \approx 1- \Sigma (\omega_1) / \omega_1$,
where $\omega_1 = \pi T$ is the first Matsubara frequency.
Another quantity that we shall obtain is $\langle n \rangle  \  vs \  \mu$
with $n$ the particle number. The slope of this curve is proportional to the 
compressibility, therefore the (Mott) insulating states will be
indicated by plateaux. 

The physics of the two band degenerate Hubbard model has been
recently considered within the dynamical mean field approach \cite{twoband,kh}.
One of the main results is that, within the paramagnetic state, 
the phase diagram shows
lines of Mott insulating states at integer fillings for
values of the interaction $U > U_c(n)$ and low enough temperatures. 
As these lines are approached as a function of filling,
a divergency in the renormalized mass is observed which
signals a correlated metallic state with an effective Fermi
energy which vanishes as $(m^* / m)^{-1}$. 
These features are reminiscent of the solution of the single
band Hubbard model which is known in quite detail \cite{mit,review}.
The key observation that we shall demonstrate in this paper is that 
the two band model with $n \leq 1$ and moderate $U$
is near a MIT line
even as it goes into a saturated
FM state. Thus, irrespective of the magnetization
the system always remains in a correlated
metallic state with a small effective Fermi energy.
The underlaying reason is simple, as the electrons become
fully polarized one may map the two band model into a {\it single
band} Hubbard model where the usual role of the spin indices
is played by the orbital ones. In other words, under a strong
magnetic field the operators carring a, say, $\downarrow$ spin 
disappear from the Hamiltonian (\ref{hamil}).

In figure 1 we show $m^* / m$ as a function of
the number of particles and
different magnetic 
fields \cite{note}. The most striking feature is how, for any field,
the mass renormalization
maintains its divergent behavior when $\langle n \rangle \rightarrow 1$ 
from below. As we argued above this occurs because the system crossover
{}from two- to one-band behavior. In particular note that for the
highest $h_{loc}$ the $m^* / m$ plot shows symmetry around $<n>=1$ as
is the case in a single band model. 
When the system becomes fully polarized, the model maps exactly
to a one-band Hubbard model in orbital space. Its associated
impurity model becomes a single Anderson impurity model with
orbital indexes playing an analogous role as the usual spin.
Thus, a small energy scale is dynamically generated as consequence
of an ``orbital Kondo'' effect \cite{notezang}
Our results also predict that the 
compounds with ${\rm x} \gtrsim 0.175$
should have an $m^*/m \lesssim 3$ which is consistent with the
relative small enhancement observed in experiments on
${\rm La_{1-x}Sr_xMnO_3}$ \cite{tokura2}.

In figure 2 we show the number of particles $\langle n \rangle$ as a function
of $\mu$ for different local magnetic fields. We observe
that plateaux are always present for fillings $\langle n \rangle = 1$ and $2$
even in the case of a strong $h_{loc}$. Since the slope of the curves is
proportional to the compressibility, the system becomes an insulator at those
fillings. In the case of $\langle n \rangle = 1$ we argued before that
under a strong local magnetic field the 
Hamiltonian maps onto a half-filled para-orbital single band model, 
therefore, the insulating state
corresponds to a Mott-Hubbard insulator for all $h_{loc}$. 
However, the character of the insulating state at $\langle n \rangle = 2$
strongly depends on $h_{loc}$. For $h_{loc} = 0$ the state is a Mott insulator
since the bands are both half-filled. On the other hand, in the polarized
state the insulator should be better thought of as a band insulator 
since the bands can
accommodate only one electron each.

The previous discussion relayed heavily on the assumption that
for $h_{loc}=1$ the system is fully polarized. Thus, in
figure 3 we plot the relative magnetic moment 
$\langle n_\uparrow - n_\downarrow \rangle / 
\langle n_\uparrow + n_\downarrow \rangle  \ vs \ \langle n \rangle$ 
at different magnetic fields in
order to check the validity of the assumption.
We observe that at zero field there is no magnetic moment as expected
while for the largest field the magnetic moment is close to unity
which indicates that all the electrons are almost
fully polarized.

It is interesting to comment on the behavior of the 
magnetization in the intermediate case \cite{note2} which illuminates aspects
of the competition between coherence and magnetization. 
At small fillings the correlation effects
due to the on-site repulsion are not important and the magnetic moment
relative to the particle number is rather
small. As the particle occupation increases
the correlation effects become 
more important (the effective mass increases) and the
magnetic moment grows rapidly due to the enhanced susceptibility of the
correlated metal. This almost saturated state persist up to filling one
and is surprising to observe that this dramatic change in the magnetization
has almost no noticeable effect on either $m^*/m$ nor the compressibility 
(figures 1 and 2). The orbital degrees of 
freedom are now playing a crucial role in order
to maintain the correlated metallic state.
As we fill the system further, the associated impurity model goes into 
a mixed valence
state and the enhanced charge fluctuations have the effect of lowering the
magnetic susceptibility. 
Finally, approaching $\langle n \rangle = 2$ the repulsive interaction
renders the electrons almost localized in a Mott state and the polarization
grows again due to the large susceptibility of the almost free moments.

To conclude we have introduced a model that contains
realistic features of the perovskite manganase oxides, namely, 
band degeneracy and
strong electronic correlations. We demonstrated that for
parameters which are appropriate for the CMR compounds
the system remains in a correlated metallic state and close to
a MIT as it goes into the saturated FM phase at low temperatures. 
The low energy physics
can then be identified with that of a single band Hubbard model
close to a MIT. We argued that, in analogy to other perovskite
TMO, this may allow for the qualitative interpretation
of a variety experiments which suggest the existence of a small enegy scale. 
We identify this energy scale with the 
renormalized Fermi energy of the coherent quasiparticle peak that
characterizes the proximity to the MIT in the dynamical mean field theory
of the Hubbard model.

Finally, is interesting to observe that in our proposed scenario
the orbital degrees of 
freedom are playing a crucial role. This seems to be 
emerging as a generic feature of correlated electron systems
which contain quasi-degenerate bands as was recently demonstrated in
the neutron scattering experiments on the classical transition metal oxide
${\rm V_2O_3}$ \cite{bao} and also in ${\rm YTiO_3}$ \cite{akimitsu}.
The application of similar techniques on the CMR compounds may serve as
an experimental test for the validity of the scenario proposed in this work.

Discussions with M. Altarelli and P. G. Radaelli are gratefully
acknowledged.

\begin{figure}
\epsfxsize=3.5in
\epsffile{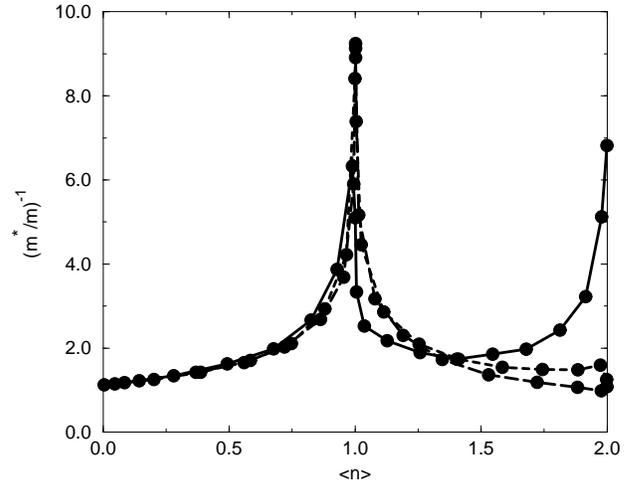}
\caption{Renormalized mass as a function of the particle occupation for
$U=3$, $T=1/8$ and $h_{loc}=0,\ 0.5,\ 1$ (full, dashed, long dashed). 
}
\end{figure}

\begin{figure}
\epsfxsize=3.5in
\epsffile{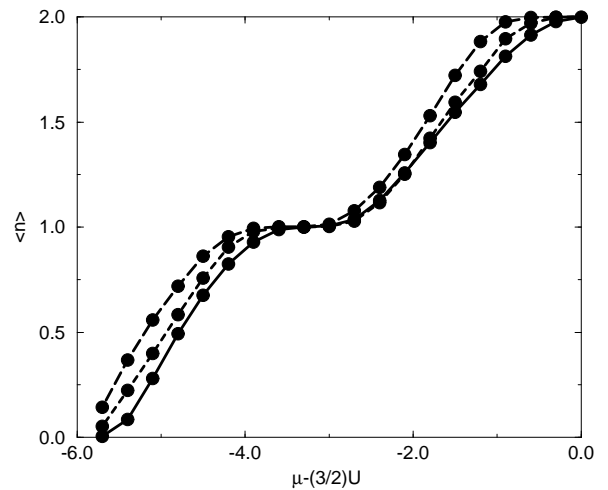}
\caption{Particle occupation as a function of the chemical potential for
$U=3$, $T=1/8$ and $h_{loc}=0,\ 0.5,\ 1$ (full, dashed, long dashed). 
}
\end{figure}

\begin{figure}
\epsfxsize=3.5in
\epsffile{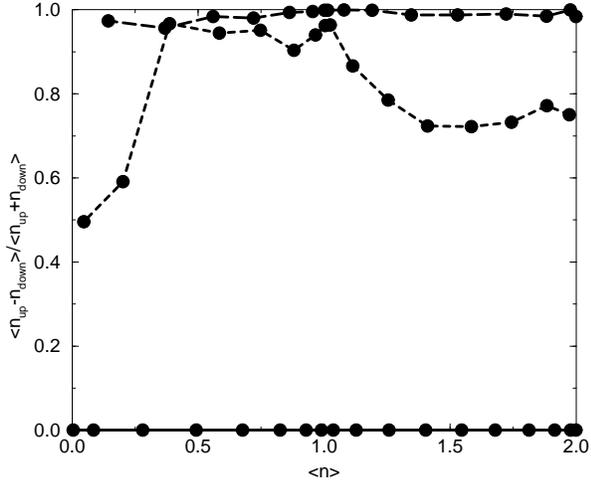}
\caption{Relative magnetic moment of the conduction electrons 
as a function of the particle number for
$U=3$, $T=1/8$ and $h_{loc}=0,\ 0.5,\ 1$ (full, dashed, long dashed). 
}
\end{figure}


\begin{thebibliography}{99}
\bibitem{note0}
We used 32 time slices
of width $\Delta\tau = 0.5$ and performed 20,000 sweeps.
Self-consistency was typically attained after 8 iterations.
Some (-) signs were observed for small occupations 
and high magnetic fields.
\bibitem{note}
Note that $\langle n \rangle = 2$ is the particle-hole symmetry point and that
fillings from $2$ to $4$ are obtained by reflection.
\bibitem{note2}
Due to the enhanced statistical noise 
in the intermediate case the numerical results
are correct within $10\%$, nevertheless, this is sufficient for the sake of
the discussion.
\end{thebibliography}
\end{document}